\documentclass[article]{gwp80}
\usepackage{graphicx}
\usepackage{amssymb}
\tabletypesize{\normalsize}  

\def\plotone#1{\centering \leavevmode
\includegraphics[width=.95\columnwidth]{#1}}

\shortauthors{Branimir Sesar}
\shorttitle{Mapping the Galactic Halo with RR Lyrae Stars}

\begin{document}
\large    
\pagenumbering{arabic}
\setcounter{page}{1}

\title{Mapping the Galactic Halo with SDSS, \\
LINEAR and PTF RR Lyrae Stars}

\author{{\noindent Branimir Sesar{$^{\rm 1}$}\\
\\
{\it (1) Division of Physics, Mathematics and Astronomy, California Institute of
Technology, Pasadena, CA 91125, USA} 
}
}

\email{(1) bsesar@astro.caltech.edu}

\begin{abstract}
We present an analysis of Galactic halo structure and substructure traced by
$\sim500$ RR Lyrae stars in the SDSS stripe 82 region. The main result of this
study is a 2D map of the Galactic halo that reaches distances of 100 kpc and
traces previously known and new halo substructures, such as the Sagittarius and
Pisces tidal streams. We also present strong direct evidence, based on both RR
Lyrae and main sequence stars, that the halo stellar number density profile
significantly steepens beyond 30 kpc from the Galactic center. Using a novel
photometric metallicity method that simultaneously combines data for RR Lyrae
and main sequence stars, we show that the median metallicity of the Sagittarius
trailing stream in the SDSS stripe 82 region is ${\rm [Fe/H]}=-1.2 \pm 0.1$ dex.
In addition to these results, we will present ongoing work on a 3D map of the
Galactic halo constructed using a sample of $\sim4000$ RR Lyrae stars that
covers 8500 deg$^2$ of northern sky and probes up to 30 kpc from the Sun.
\end{abstract}

\section{Galactic Halo as the Rosetta Stone for Galaxy Formation}

Studies of the Galactic halo can help constrain the formation history of the
Milky Way and the galaxy formation process in general. For example,
state-of-the-art simulations of galaxy formation predict numerous substructures,
such as tidal tails and streams, in halos of Milky Way-size galaxies (e.g., see
Fig.~14 in \citealt{joh08} or Fig.~6 in \citealt{coo10}). The amount,
morphology, kinematics, and chemical composition of these substructures depend
on the accretion history of the simulated galaxy \citep{joh08, coo10}.
Therefore, if we map the substructures in the Galactic halo and compare the
resultant maps with simulations, we will be able to constrain the formation
history of the Milky Way.

Maps of the Galactic halo that probe galactocentric distances ($R_{\rm GC}$) of
$\sim20-30$ kpc have already been made using near turn-off main sequence stars
\citep{jur08, bel08} selected from the Sloan Digital Sky Survey (SDSS;
\citealt{yor00}). Unfortunately, the largest discrepancy between simulated halos
with different formation histories occurs at distances beyond 30 kpc (e.g.,
compare panels in \citealt{coo10} Fig.~6). Therefore, to discriminate between
different formation histories we need to map the Galactic halo beyond 30 kpc and
to do that we need to use tracers that are brighter than main-sequence turnoff
stars.

\section{Mapping the Galactic Halo with SDSS Stripe 82 RR Lyrae Stars}

To map the Galactic halo beyond 30 kpc, we used RR Lyrae stars selected from the
SDSS stripe 82 region ($20^h 32^m <\alpha_{J2000.0} < 04^h 00^m$,
$-1.26\arcdeg<\delta_{J2000.0} < +1.26\arcdeg$, $\sim280$ deg$^2$). This region
has been repeatedly observed during SDSS-I and SDSS-II parts of the survey and
on average, there are about 30 observations per source. The photometric
precision of SDSS Stripe 82 data is $\sim0.02$ mag at the bright end
($r < 18$) and $\sim0.05$ mag at $r\sim21$ \citep{ses07}. For comparison, the
apparent magnitude of an RR Lyrae star at 100 kpc is $r\sim21$.

The candidate RR Lyrae stars were pre-selected using SDSS colors and using some
low-level variability statistics, such as rms scatter and chi-square per degree
of freedom (for details on the selection algorithm see \citealt{ses10a}). The
period of variability was found for each candidate using an implementation of
the {\em Supersmoother} algorithm \citep{rei94}, the candidate light curve was
folded using that period and a custom set of ugriz template light curves was
fitted to the period-folded light curve for identification. In total, we have
identified about 400 type-$ab$ (RRab) and about 100 type-$c$ (RRc) RR Lyrae
stars in the SDSS stripe 82 region.

To calculate distances for RRab stars we used the \citet{chaboyer99}
$M_V-[Fe/H]$ relation
\begin{equation}
M_V = (0.23\pm0.04)[Fe/H] + (0.93\pm0.12)\label{abs_mag}
\end{equation}
and assumed ${\rm [Fe/H]}=-1.5$ dex (median halo metallicity; \citealt{ive08})
as the metallicity of all RR Lyrae stars in our sample\footnote{Spectroscopic
metallicities were not available for the majority of stars in our sample.}. This
relation gives $M_{\rm V}=0.6$ for the absolute magnitude of RRab stars in the
Johnson $V$ band. The resulting fractional uncertainty in distance due to
unknown metallicity, evolutionary effects, and photometry is $\sim5\%$. RRc
stars were not used in the rest of this study.

\begin{figure*}
\centering
\plotone{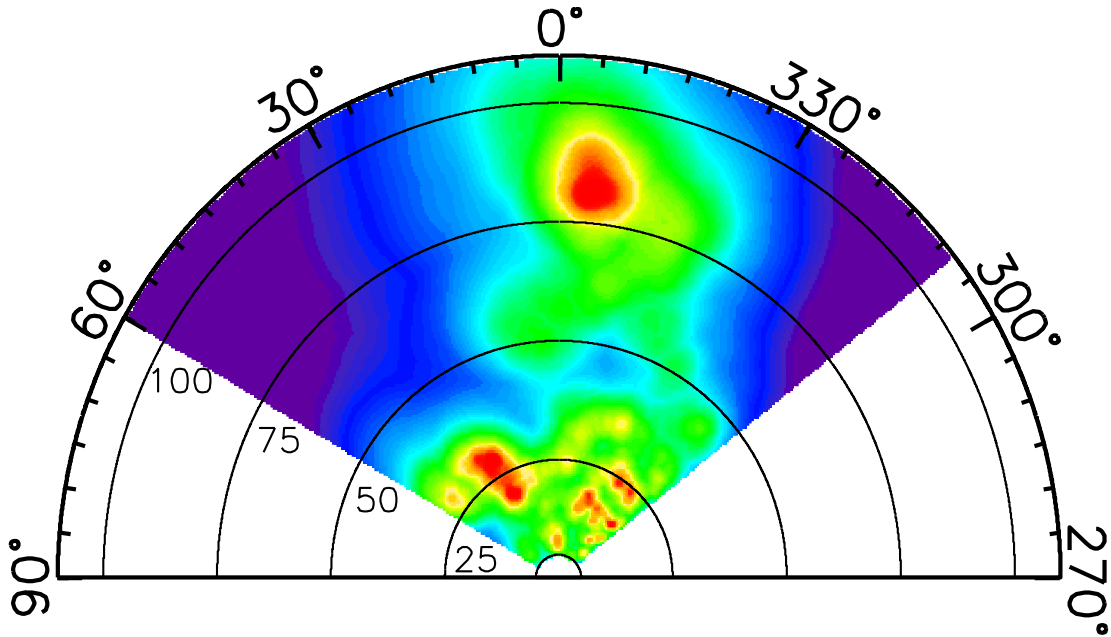}
\vskip0pt
\caption{
Halo substructures traced by RR Lyrae stars in SDSS stripe 82. The heliocentric
distance in kpc is in the radial direction and angle indicates equatorial right
ascension. The red color indicates substructures, green color shows the smooth
halo, and the blue color shows regions where there are almost no stars. The
Pisces Overdensity/Stream is located $\sim80$ kpc from the Sun at
${\rm R.A.}\sim355$ deg, the trailing arm of the Sagittarius tidal stream
\citep{ive03} is passing through the stripe 82 plane at (R.A., helio.
dist.)$=$($\sim30$ deg, 25 kpc), and the Hercules-Aquila Cloud is located at
($\sim330$ deg, $\lesssim25$ kpc). Figure adapted from \citet{ses10a} Fig.~11.
\label{fig1}
}
\end{figure*}

A Bayesian number density estimator \citep{ive05} was applied on the spatial
distribution of RR Lyrae stars and the calculated number densities of RR Lyrae
stars were compared to values predicted by a smooth, oblate halo model with a
density power-law slope of -2.7 (best-fit halo model from \citealt{jur08}).
Fig.~\ref{fig1} shows the comparison of observed and model-predicted number
densities on a logarithmic scale. The green regions are in agreement with the
smooth halo model, red regions are overdense by a factor of 10, and blue regions
are underdense by a factor of 10 compared to the model.

Within about 30 kpc, the observed halo follows the oblate power-law model. Two
substructures are clearly visible: the Sgr trailing arm and the Hercules-Aquila
Cloud \citep{bel07}. Beyond 30 kpc, the model predicts more stars than what is
actually observed, indicating that the observed number density profile steepens.
This steepening is also present when main-sequence stars are used as tracers,
as shown in \citet{sji11} Figs.~9 and~10 and in Fig.~\ref{fig2} below.
\begin{figure*}
\centering
\plotone{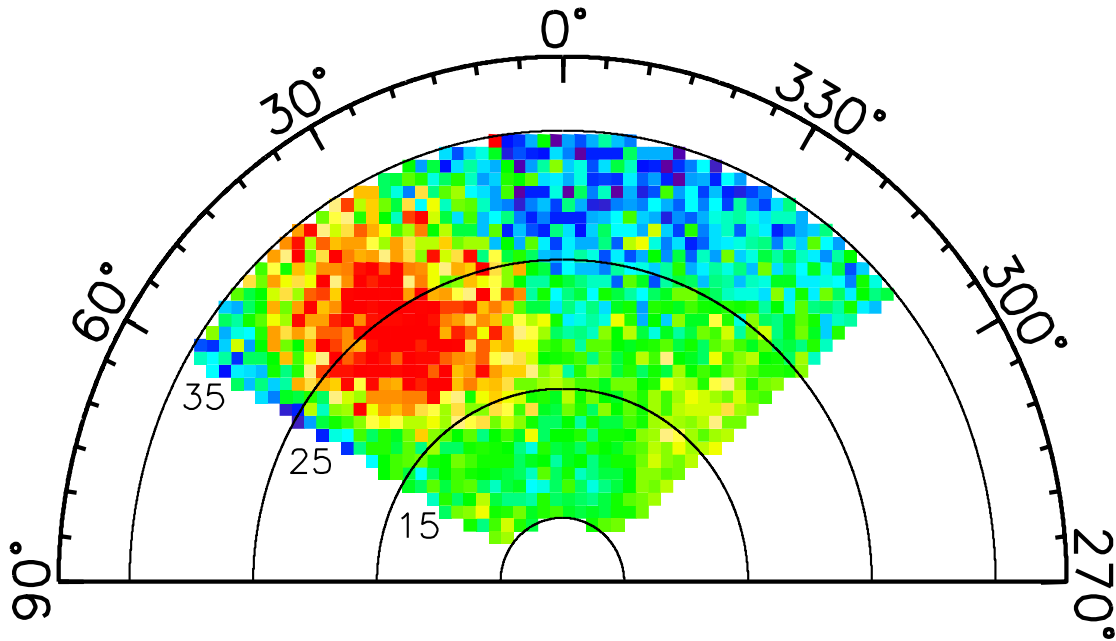}
\vskip0pt
\caption{
Halo substructures traced by near turnoff main-sequence stars in SDSS stripe 82.
The distance limit in this map is at 40 kpc from the Sun. Note the Sagittarius
tidal stream, Hercules-Aquila Cloud, and the steepening of the halo density
profile beyond $\sim30$ kpc. Figure adapted from \citet{ses10a} Fig.~24.
\label{fig2}
}
\end{figure*}

The overdensity seen in Fig.~\ref{fig1} at about 80 kpc was first reported in
\citet{ses07} (``J'' clump) and was later renamed the Pisces overdensity by
\citet{wat09}. \citet{kol09} and \citet{ses10b} have observed this substructure
spectroscopically and have detected two velocity peaks, suggesting that the
substructure may be a tidally disrupted dwarf galaxy. A more extended view of
this substructure was provided by \citet{sha10} who used M-giants selected from
2MASS.

\subsection{A New Photometric Metallicity Method}

As shown in Figs.~\ref{fig1} and~\ref{fig2}, we have detected the Sagittarius
dSph tidal stream (trailing arm) in SDSS stripe 82 as an overdensity of RR Lyrae
and main-sequence stars. These detections have allowed us to develop and test a
new photometric metallicity method that can be used when SDSS $u$ band
observations are not available.

\begin{figure*}
\centering
\plotone{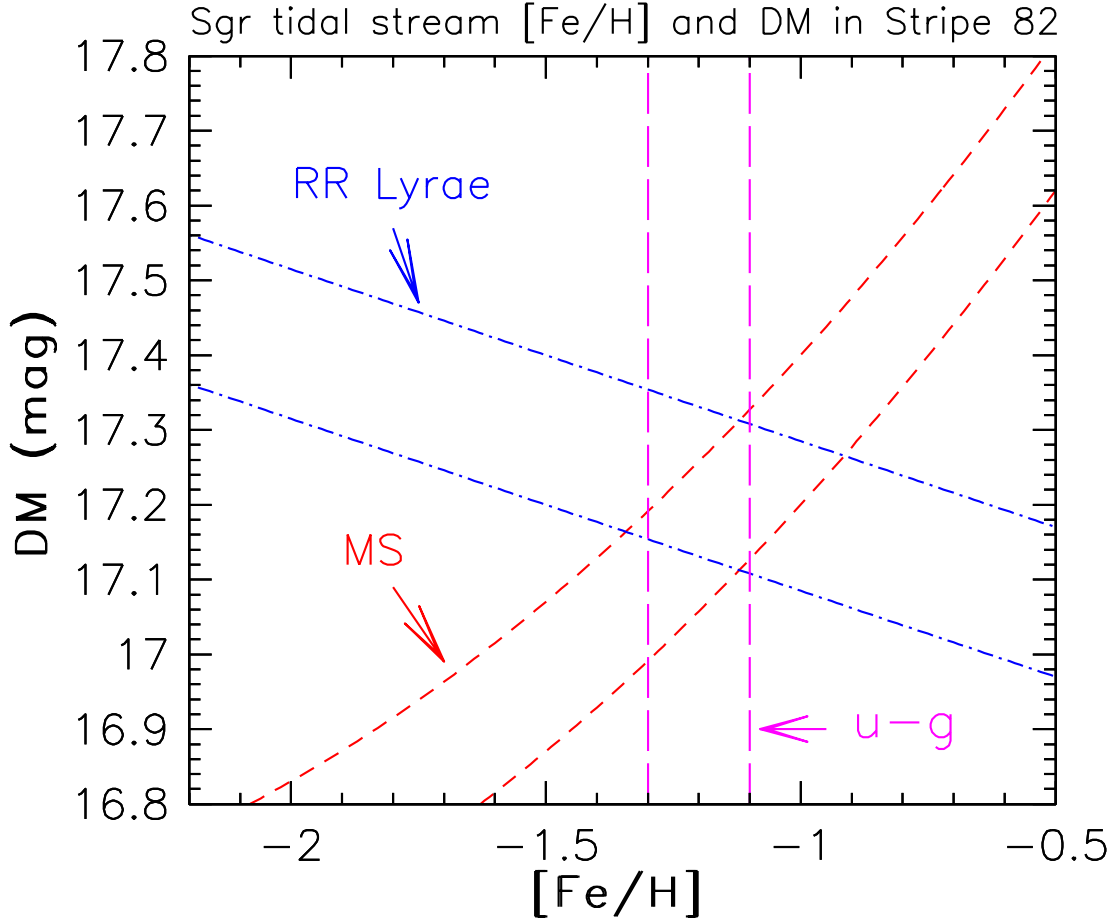}
\vskip0pt
\caption{
A summary of the constraints on the distance and metallicity of the Sagittarius
dSph tidal stream (trailing arm). The dot-dashed lines show a constraint
obtained from the mode of the apparent magnitude distribution of RR Lyrae stars,
with $\pm0.1$ mag adopted as the uncertainty. Diagonal short-dashed lines are
analogous constraints obtained from the median apparent magnitude of
main-sequence stars with $0.4 < g - i < 0.5$. The vertical longdashed lines mark
the median photometric metallicity for main sequence stars obtained using the
\citet{ive08} photometric metallicity method, with $\pm0.1$ dex adopted as the
uncertainty. All three constraints agree if the distance modulus is $DM = 17.2$
mag and ${\rm [Fe/H]} = -1.2$ dex.
\label{fig3}
}
\end{figure*}

As shown in Fig.~\ref{fig3}, for fixed apparent magnitudes, the distance modulus
of RR Lyrae and main-sequence stars changes with metallicity, but with opposite
signs. For RR Lyrae stars, the distance modulus decreases with increasing
metallicity, as dot-dashed lines show. The distance modulus of main-sequence
stars, on the other hand, increases with increasing metallicity, as dashed lines
show. If we assume that a clump of main-sequence stars belongs to the same
substructure as the clump of RR Lyrae stars, we can solve for metallicity and
distance modulus. In the case of Sagittarius trailing arm this method gives
median metallicity of about -1.2 dex, and agrees well with the median
spectroscopic metallicity obtained by \citet{car10} (${\rm [Fe/H]=-1.15}$ dex).

The $u-g$ color of main-sequence stars can also be used as an estimator of
metallicity, as detailed in \citet{ive08}. Using this method and main-sequence
stars in SDSS stripe 82, we estimated $-1.2$ dex as the metallicity of the
Sagittarius trailing arm. However, the Ivezi\'c et al. photometric metallicity
method requires $u-g$ observations of main-sequence stars. These observations
are difficult to obtain and some current and upcoming sky surveys will not have
$u$-band detections at all. The advantage of the above method is that it does
not need $u$-band observations, and therefore it can be used by surveys such as
PTF, PanSTARRS and DES that will not have observations in the $u$ band.

\vfill\eject

\section{Mapping the Galactic Halo with LINEAR and PTF RR Lyrae Stars}

The biggest disadvantage of SDSS stripe 82 data is its small sky coverage
(only $\sim1\%$ of the sky), meaning that some of the results and conclusions
presented here may not be representative of the entire halo. To address this
issue, we have selected about 4000 RRab stars from the LINEAR survey (covering
10,000 deg$^2$ of sky) and have started to map the Galactic halo in 3D in order
to verify the above results based on SDSS stripe 82 data with much smaller
sky coverage ($\sim300$ deg$^2$).

LINEAR stands for the Lincoln Near-Earth Asteroid Research. This ongoing survey
\citep{sto00} has been observing the northern sky since 1998 and is a premier
source for variability studies. Unfortunately, detections of asteroids were more
important to this survey than photometry so the quality of original data is not
as good as in SDSS stripe 82. Following a procedure similar to the one described
in \citet{ses06}, we have recalibrated LINEAR photometry using SDSS data and
have created a database that contains 8.2 million objects with more than 200
observations per object. A sample of about 4000 RRab stars has been selected
from this dataset using the selection algorithm from \citet{ses10a}. This sample
is 90\% complete up to 30 kpc, covers 10,000 deg$^2$ of sky, and the
contamination (fraction of non-RR Lyrae stars) is less than 2\%.
\begin{figure*}
\centering
\plotone{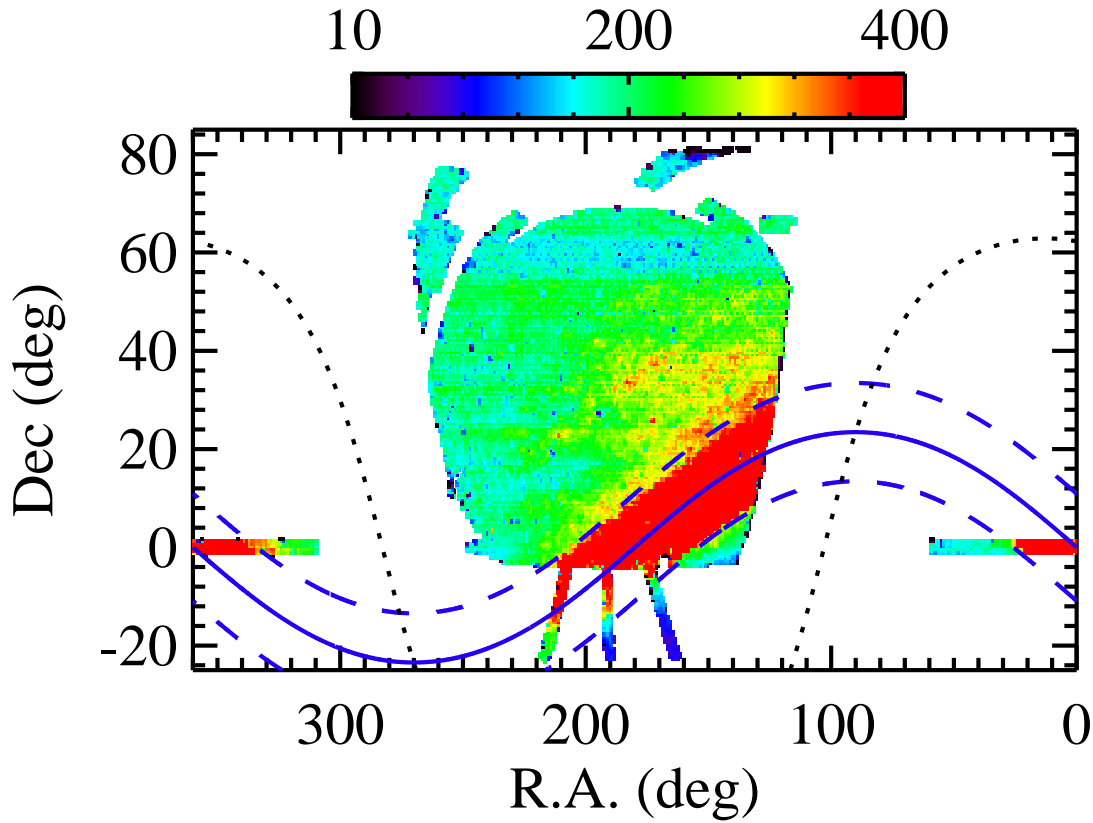}
\vskip0pt
\caption{
The median number of observations per object in the recalibrated LINEAR dataset
as a function of equatorial J2000.0 right ascension and declination coordinates.
The values are color-coded according to the legend, with values outside the
range saturating. The dashed lines show $\pm10\arcdeg$ of the ecliptic plane
({\em solid line}) and the Galactic plane is shown as a dotted line. The average
number of observations per object within $\pm10\arcdeg$ of the ecliptic plane is
$\sim460$, and $\sim200$ elsewhere.
\label{fig4}
}
\end{figure*}

\begin{figure*}
\centering
\plotone{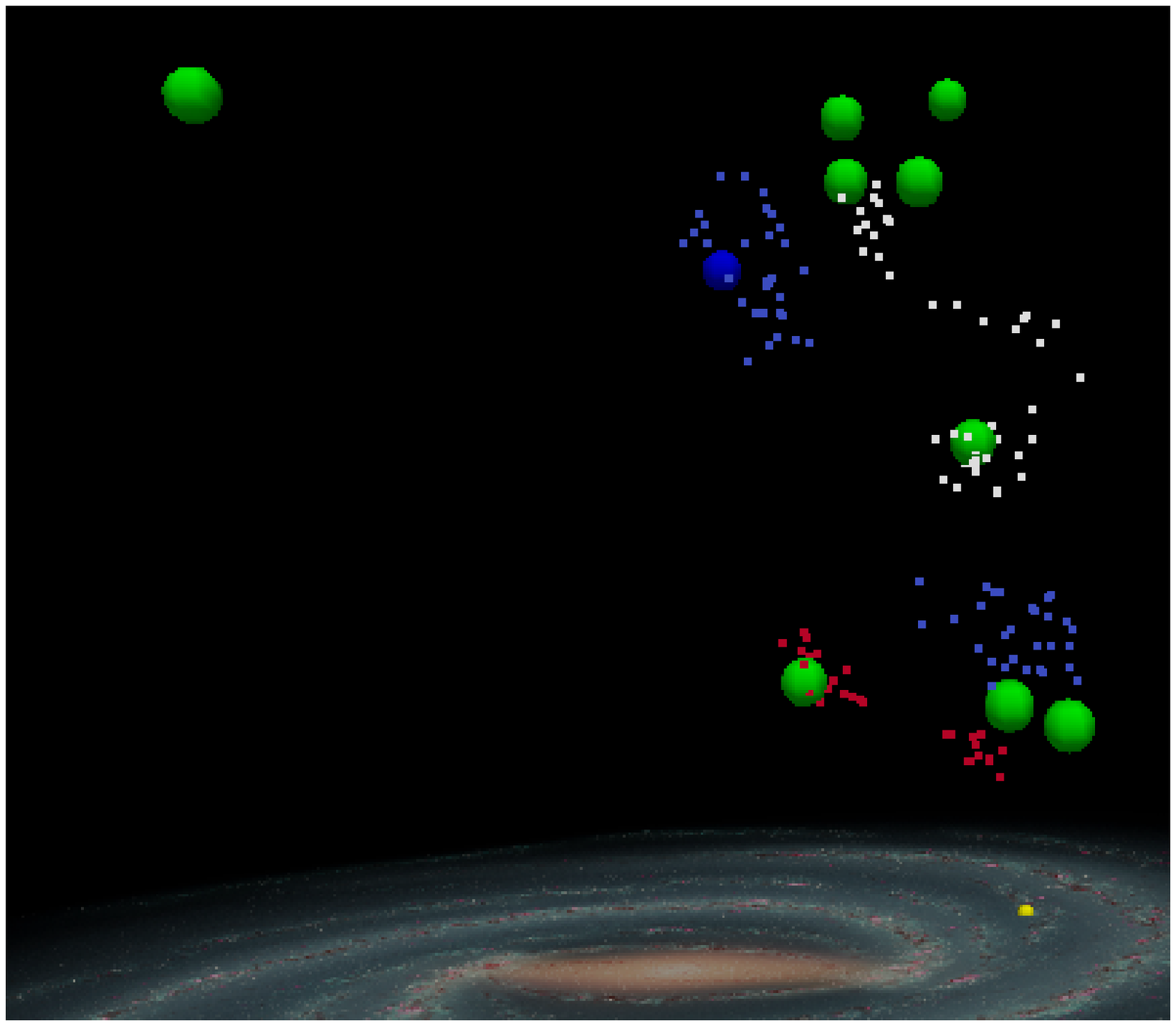}
\vskip0pt
\caption{
A 3D rendering of halo substructures detected using LINEAR RRab stars. The
points show RRab stars clustered by Enlink into 6 significant groups (there are
two red, blue, and white clusters). The Sun's position in the Galactic disk is
indicated by a yellow sphere. The green spheres show positions of some globular 
clusters and the blue sphere shows the position of the Virgo Stellar Stream
according to \citet{duf06}. The spatial extent of some groups suggests that
they may be tidal streams.
\label{fig5}
}
\end{figure*}

In order to detect substructures in the region of halo probed by LINEAR RR Lyrae
stars, we have applied the {\em Enlink} group-finding algorithm \citep{sj09} to
our sample of LINEAR RRab stars. The Enlink algorithm has identifed 6
significant halo groups based only on positions of RR Lyrae stars, and we show
these groups in Fig.~\ref{fig5}. Even though most of these groups can be
associated with globular clusters or known halo streams (e.g., Virgo Stellar
Stream; \citealt{duf06}), the morphology of some groups seems to indicate that
they are tidal streams, possibly originating from some of the globular clusters.

Even though the LINEAR survey covers a much wider area of the sky than SDSS
stripe 82, it is quite shallow ($r<18$) and RR Lyrae selected from it probe only
the inner part of the Galactic halo (within 30 kpc from the Sun). What is needed
is a multi-epoch survey that covers a wide area of the sky and is deep at the
same time. The survey that satisfies these criteria and is the Palomar Transient
Factory.

PTF is a wide-area, two-band (SDSS-g$^\prime$ and Mould-R filters), deep
($R\sim21$, single-epoch; $R\sim23$, co-added) survey aimed at systematic
exploration of the optical transient sky. It will provide accurate photometry
(with systematic uncertainties $\lesssim0.02$ mag) for more than 10,000 deg$^2$
of sky in two cadences: i) 5DC $-$ two 60s exposures separated by one hour and
repeated after 5 days, and ii ) DyC $-$ a dynamical cadence designed to explore
transient phenomena on timescales longer than 1 min and shorter than 5 days.
These are optimal cadences for the observation of RR Lyrae stars (period of
pulsation $\sim0.6$ days) and will ensure a dense phase coverage of light
curves. This survey is ongoing and will provide 60 epochs by the end of the
survey, enough for an efficient selection of RR Lyrae stars.

\section{Conclusions}

We have presented the most complete sample of RR Lyrae stars identified so far
in the SDSS stripe 82 data set, consisting of 379 RRab and 104 RRc stars. Our
visual inspection of single-band and color light curves insures that the sample
contamination is essentially negligible. Although the sky area is relatively
small compared to other recent surveys, such as \citet{kel08} and \citet{mic08},
this RR Lyrae sample has the largest distance limit to date ($\sim$100 kpc).

The high level of completeness and low contamination of our resulting sample, as
well as the more precise distance estimates, enabled a more robust study of halo
substructure than was possible in our first study. Our main result remains: the
spatial distribution of halo RR Lyrae at galactocentric distances 5--100 kpc is
highly inhomogeneous. A comparison of the observed spatial distribution of RR
Lyrae stars and main sequence stars to the \citet{jur08} model, which was
constrained by main sequence stars at distances up to 20 kpc, strongly suggests
that the halo stellar number density profile steepens beyond $\sim$30 kpc. While
various indirect evidence for this behavior, based on kinematics of field stars,
globular clusters, and other tracers has been published (\citealt{car07}; and
references therein), our samples provide a direct measurement of the stellar
halo spatial profile beyond the galactocentric distance limit of $\sim$30 kpc. A
similar steepening of the spatial profile was detected using candidate RR Lyrae
stars by \citet{kel08}.

We introduced a novel method for estimating metallicity that is based on the
absolute magnitude vs.~metallicity relations for RR Lyrae stars and main
sequence stars (calibrated using globular clusters). This method does not
require the $u$-band photometry, and will be useful to estimate metallicity of
spatially coherent structures that may be discovered by the Dark Energy and
Pan-STARRS surveys. While the existing SDSS data are too shallow to apply this
method to the Pisces stream, we used it to obtain a metallicity estimate for the
Sgr tidal stream that is consistent with an independent estimate based on the
photometric $u$-band method for main sequence stars. Our result, $[Fe/H]=-1.2$,
with an uncertainty of $\sim$0.1 dex, strongly rules out the hypothesis that the
trailing arm of the Sgr dSph tidal stream has the same metallicity as halo field
stars ($[Fe/H]=-1.5$), as suggested by \citet{wat09}. \citet{yan09} detected
peaks at $[Fe/H]=-1.3$ and at $[Fe/H]=-1.6$ using SDSS spectroscopic
metallicities for blue horizontal-branch (BHB) stars from the trailing arm.
However, both the Yanny et al.~and Watkins et al.~results could be affected by
systematic errors in the SDSS metallicity scale for BHB stars. Our results
suggest that the SDSS metallicity scale for BHB stars could be biased low by
about 0.3 dex (relative to the SDSS metallicity scale for main sequence stars,
and assuming that RR Lyrae and main sequence stars from the Sgr tidal stream in
stripe 82 area have the same metallicity distributions).

Simulations by \citet{bj05} and \citet{joh08} predict that there should be a
difference in the chemical composition between stars in the inner halo that was
built from accretion of massive satellites about 9 Gyr ago, and outer halo
dominated by stars coming from dSph satellites that were accreted in the last 5
Gyr (see Fig.~11 in \citealt{bj05}). These accreted dSph satellites were
presumably more metal-poor than the massive satellites accreted in earlier
epochs (see Fig.~3 in \citealt{rob05}). Other recent simulation studies support
these conclusions. For example, \citet{dh08} find evidence in their simulation
for a strong concentration of (relatively) higher metallicity stars at distances
close to the Galactic center, and the presence of (relatively) lower metallicity
stars at distances beyond 20 kpc from the center. \citet{zol09} find from their
simulations that their inner halos include stars from both in-situ formed stars
and accreted populations, while their outer halos appear to originate through
pure accretion and disruption of satellites.

Simulations indicate that high surface brightness substructures in the halo
originate from single satellites, typically massive dSph which tend to be
accreted over the last few Gyr \citep{bj05}, and these massive galaxies are
expected to be more more metal-rich than halo field stars \citep{fon08}. The
results from \citet{ive08} and the results presented here seem to support this
prediction. The inner halo has a median metallicity of $[Fe/H]=-1.5$, while at
least two strong overdensities have higher metallicities -- the Monoceros stream
has $[Fe/H]=-1.0$, and for the trailing part of the Sgr tidal stream we find
$[Fe/H]=-1.2$. We emphasize that these three measurements are obtained using the
same method/calibration and the same data set, and thus the measurements of
relative differences are expected to be robust.

Our result that (inner-) halo stellar number density profile steepens beyond
$\sim$30 kpc is limited by the relatively small distance limit for main sequence
stars (35 kpc), the sparseness of the RR Lyrae sample ($\sim$500 objects), and
the small survey area ($\sim$300 deg$^2$). Ideally, the halo stellar number
density profile should be studied using numerous main sequence stars detected
over a large fraction of sky. To do so to a distance limit of 100 kpc, imaging
in at least $g$ and $r$ bands (or their equivalent) to a depth several
magnitudes fainter than the co-added SDSS stripe 82 data is required ($r>25$).
Pan-STARRS, the Dark Energy Survey and LSST are planning to obtain such data
over large areas of sky. The LSST, with its deep $u$-band data, will also extend
metallicity mapping of field main sequence stars over half of the sky in the
south; see \citet{ive08} for details. For substructures to be potentially
discovered in the north by Pan-STARRS, the method presented here can be used to
estimate the metallicity of spatially coherent structures even without the
$u$-band data.


\begin{thebibliography}

\bibitem[Bell et al.(2008)]{bel08}Bell, E. F. et al. 2008, \apj, 680, 295
\bibitem[Belokurov et al.(2007)]{bel07}Belokurov, V. et al. 2007, \apjl, 657, 89
\bibitem[Bullock \& Johnston(2005)]{bj05}Bullock, J. S. \& Johnston, K. V. 2005,
\apj, 635, 931
\bibitem[Carlin et al.(2010)]{car10}Carlin, J. L. et al. 2010, \baas, 42, 319
\bibitem[Carollo et al.(2007)]{car07}Carollo, D. et al. 2007, \nat, 450, 1020
\bibitem[Chaboyer(1999)]{chaboyer99}Chaboyer, B. 1999, in ``Post-Hipparcos
cosmic candles'', Eds. A. Heck \& F. Caputo, Kluwer Academic Publishers, p. 111
\bibitem[Cooper et al.(2010)]{coo10}Cooper, A. P. et al. 2010, \mnras, 406, 744
\bibitem[De Lucia \& Helmi(2008)]{dh08}De Lucia, G. \& Helmi, A. 2008, \mnras,
391, 14
\bibitem[Duffau et al.(2006)]{duf06}Duffau, S. et al. 2006, \apjl, 636, 97
\bibitem[Font et al.(2008)]{fon08}Font, A. S. et al. 2008, \apj, 673, 215
\bibitem[Ivezi\'{c} et al.(2003)]{ive03}Ivezi\'{c}, \v{Z}. et al. 2003, \memsai,
74, 978
\bibitem[Ivezi\'{c} et al.(2005)]{ive05}Ivezi\'{c}, \v{Z}. et al. 2005, \aj,
129, 1096
\bibitem[Ivezi\'{c} et al.(2008)]{ive08}Ivezi\'{c}, \v{Z}. et al. 2008, \apj,
684, 287
\bibitem[Johnston et al.(2008)]{joh08}Johnston, K. V. et al. 2008, \apj,
689, 936
\bibitem[Juri\'c et al.(2008)]{jur08}Juri\'c, M. et al. 2008, \apj, 673, 864
\bibitem[Keller et al.(2008)]{kel08}Keller, S. C. et al. 2008, \apj, 678, 851
\bibitem[{{Kollmeier} {et~al.}(2009){Kollmeier}, {Gould}, {Shectman},
  {Thompson}, {Preston}, {Simon}, {Crane}, {Ivezi{\'c}}, \&
    {Sesar}}]{kol09}
	{Kollmeier}, J.~A. {et~al.} 2009, \apjl, 705, L158
\bibitem[Miceli et al.(2008)]{mic08}Miceli, A. et al. 2008, \apj, 678, 865
\bibitem[Rau et al.(2009)]{rau09}Rau, A. et al. 2009, \pasp, 121, 1334
\bibitem[Reimann(1994)]{rei94}Reimann, J. D. 1994, Ph.D. thesis,
Univ.~California, Berkeley
\bibitem[Robertson et al.(2005)]{rob05}Robertson, B. et al. 2005, \apj, 632, 872
\bibitem[Sesar et al.(2006)]{ses06}Sesar, B. et al. 2006, \aj, 131, 2801
\bibitem[Sesar et al.(2007)]{ses07}Sesar, B. et al. 2007, \aj, 134, 2236
\bibitem[Sesar et al.(2010a)]{ses10a}Sesar, B. et al. 2010a, \apj, 708, 717
\bibitem[Sesar et al.(2010b)]{ses10b}Sesar, B., Vivas, A. K., Duffau, S. \&
Ivezi\'c, \v{Z}. 2010b, \apj, 717, 133
\bibitem[Sesar, Juri\'c \& Ivezi\'c(2011)]{sji11}Sesar, B., Juri\'c, M. \&
Ivezi\'c, \v{Z}. 2011, \apj, 731, 4
\bibitem[Sharma \& Johnston(2009)]{sj09}Sharma, S. \& Johnston, K. 2009, \apj,
703, 1061
\bibitem[Sharma et al.(2010)]{sha10}Sharma, S. et al. 2010, \apj, 722, 750
\bibitem[Stokes et al.(2000)]{sto00}Stokes, G. et al. 2000, Icarus, 148, 21
\bibitem[{{Watkins} {et~al.}(2009){Watkins}, {Evans}, {Belokurov}, {Smith},
  {Hewett}, {Bramich}, {Gilmore}, {Irwin}, {Vidrih}, {Wyrzykowski}, \&
    {Zucker}}]{wat09}
	{Watkins}, L.~L. {et~al.} 2009, \mnras, 398, 1757
\bibitem[Yanny et al.(2009)]{yan09}Yanny, B. et al. 2009, \apj, 700, 1282
\bibitem[York et al.(2000)]{yor00}York, D. G. et al. 2000, \aj, 120, 1579
\bibitem[Zolotov et al.(2009)]{zol09}Zolotov, A. et al. 2009, \apj, 702, 1058

\end{thebibliography}
\end{document}